\documentstyle[epsf,psfig,multicol,aps,prl]{revtex}

\newcommand{\be}{\begin{equation}}
\newcommand{\ee}{\end{equation}}

\begin{document}

\title{Life in the Stockmarket - a Realistic Model for Trading}

\author{Fabio Franci, and Lorenzo Matassini\\
{\small Max-Planck-Institut f\"ur Physik komplexer Systeme,
N\"othnitzer Str.\ 38, D 01187 Dresden, Germany}}

\maketitle
\vglue2ex

\begin{abstract}
We propose a frustrated and disordered many-body model of a stockmarket in which independent adaptive traders can 
trade a stock subject to the economic law of supply and demand. We show that the typical scaling properties and the 
correlated volatility arise as a consequence of the collective behavior of agents: With their interaction they determine
a price which in turn affects their future way of investing.
We introduce only one type of investors, since they all share the same hope: They \emph{simply} want to maximize the 
profit minimizing the risk. The best utilization of resources occurs at a critical point characterized by the transition
between the excess-demand and the excess-supply phases.

\vspace*{2mm}
\end{abstract}

\begin{multicols}{2}
Financial prices have been found to exhibit some universal features that resemble the scaling laws characterizing 
physical systems in which large numbers of units interact \cite{haken,mant_nature,lux_nature,lux2,mantegna}.
Given that, one challenge is to build up a model which is able to reproduce this behavior and whose parameters
have a physical meaning.
We propose such a model, belonging to the frustrated and disordered many-body systems. Frustration enters in that not
all the individual inclinations can be satisfied simultaneously, whereas the model is disordered because traders have
randomly chosen expectations and resources but they share the same strategy.
In contrast with previuos works we do introduce only one kind of investor: On our opinion the usual distinction between 
fundamentalists and noise traders is not necessary. Looking at both 
strategies, in fact, we note that \emph{fundamentalists} follow the premise of the efficient market hypothesis 
in that they expect the price to follow the fundamental value of the asset: A fundamentalist trading strategy 
consists of buying when the actual market price is believed to be below the fundamental value and selling in 
the opposite case. \emph{Noise traders}, on the other hand, do not believe in an immediate tendency of the 
price to revert to its underlying fundamental value: They try to identify price trends and consider the 
behavior of other traders as a source of information. This gives rise to the tendency towards 
\emph{herding behavior}.

In \cite{lux_nature} a model is introduced where the main building blocks are movements of 
individuals from one group to another together with the exogenous changes of the fundamental value. We believe 
that such a requirement is somehow artificial. It is in fact clear
that without the fundamentalists the price would follow a trend at infinitum and without the noise traders
the price would never excape the range of its fundamental value, as clearly stated in \cite{bak}, where a 20\% of
fundamentalists is enough to confine the price within the range of the rational traders. 
We want to raise serious doubts on the blind use of the fundamental price, expecially when assumed that the relative 
changes are Gaussian random variables: Why should agents risk money just believing in a random walk behavior?

A more realistic assumption is based on the following wish: Gaining the maximum, taking the smallest possible risk.
Agents make decisions to buy or to sell, adjust prices, and so on according to the information available at the time,
as well as individual preferences such as tolerance for risk and time deadlines.
We therefore simulate the book for the ask and for the bid. Every trader, when buying a share, has to identify a fair 
price and then put an order keeping in mind a target price and a stop-loss price. These quantities are the result of the 
interaction with other agents, the study of past values of the price and the influence of incoming news.
The decision to sell some shares is made on the basis of the current price (to be compared with the personal target
and stop-loss price) and of the \emph{age} of the shares: After a sufficently long period of time
agents start to ask themselves whether it is worth keeping the money invested in that way. The model leads to a 
\emph{self-organized criticality} which is responsible of the alternation between \emph{laminar} and \emph{turbulent}
trading \cite{paczuski}.

\vspace*{0.3cm}
\centerline{\psfig{file=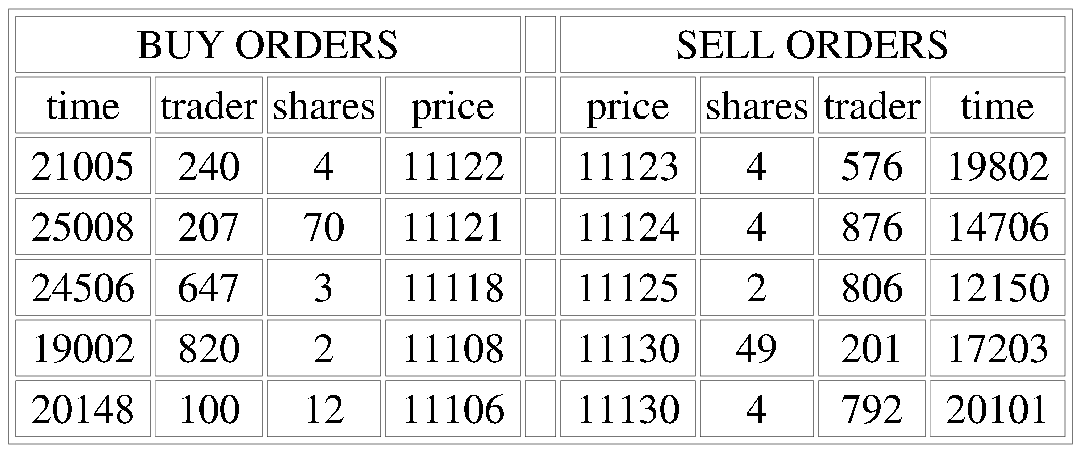,width=8cm}}
\vspace*{0.1cm}
{\small TABLE 1. Example of the first five levels of the book. No transaction can take place 
because the highest buying price is smaller than the cheapest selling order. Entries are ordered according to 
price and occurrence time in case of equal prices.}
\vspace*{0.1cm}

We consider $N$ traders and one stock with $M$ shares on the market. Each trader is characterised through 
the following information:
\emph{(i)} Initial amount of money.
\emph{(ii)} Inclination towards investment: Usually traders tend to keep cash a part of their resources, 
            in order to be able to have money to exploit the market at special time.
\emph{(iii)} Number of shares owned.
\emph{(iv)} List of friends with which he is sharing information, to model the herding effect.
\emph{(v)} Invested money, to keep trace of the average buying price.
\emph{(vi)} Desired gain.
\emph{(vii)} Maximum loss.
\emph{(viii)} Threshold: Amount of time after which the trader may start to change idea about his/her investment.

Every order is stored in the corresponding list, according to the type of it (buy or sell), to the
requested price and to the time at which it was submitted. A transaction occur whenever the cheapest price
among the sell list matches with the most expensive offer in the buyers' list: This value defines the market 
price of the stock at that particular instant and it will of course affect the future behavior of the traders.

We provide a mechanism to produce \emph{news}, whose purpose is to let all the agents know some information
about the overall behavior of the market, i.e. the unbalance of the two books and the actual volatility.
It is very important to note that these signals are endogenous: The information they provide was already 
present in the system. In this way our model takes into account both a \emph{local and global coupling}, via shared
information with neighbours and generation of news and advertisements, respectively.

\begin{figure}
\centerline{\psfig{file=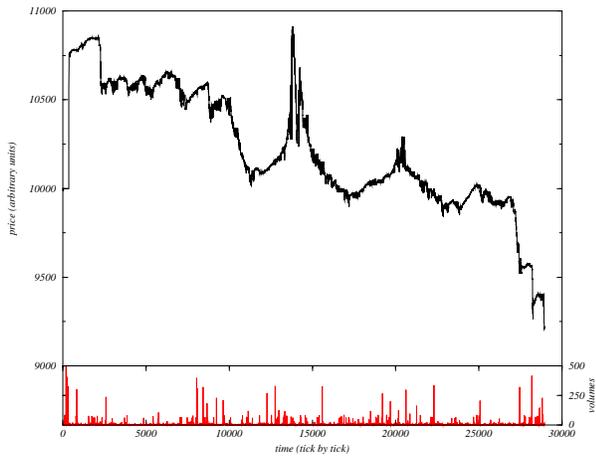,width=8cm,angle=270}}
\caption[]{\small\label{fig:serie}
Typical snapshot from a simulation run. Upper panel: Development with time of the market price.
Lower panel: Development with time of the corresponding volumes of exchange.}
\end{figure}

The simulation consists of two parts. At the beginning we assign all the shares to one trader and we 
broadcast advertisements to induce people to put a buy order. The purpose of this first step is to simulate
the {\bf Initial Public Offer} ({\bf IPO}) and the selected trader can be thought as the bank responsible
of the initial selling of the stock. This part ends when the number of shares owned by the IPO trader vanishes. 
We then reset his amount of money and enter the second phase.

At each update step the algorithm performs the following operations:
\emph{(i)}   Select randomly a trader.
\emph{(ii)}  If the trader has no pending order and no share then he/she is eventually willing to
             buy, formulating a limit price, a target price and a stop-loss price and inserting a buying order.
\emph{(iii)} In case of owning some shares, the trader may decide to sell, according to the
             market price and the threshold.	  
\emph{(iv)}	 If the trader has a pending order, he/she may change some parameter because the conditions
             that led to that decision may have changed.
We suppose that every trader can afford only one pending order.	Every time a buyer has got all the desired 
shares, he/she is immediately asked whether he/she wants to place a selling order.

When formulating the prices, every trader makes the decision in a \emph{deterministic} way, computing a weighted
average among the opinion of some acquaintance, the indication of the news and some past values of the stock price.
Every time that two complementary orders match, we define the market price for that particular instant:
Usually only one of the two orders disappears, namely the one with the smallest amount of
shares involved. The other cannot be removed from the corresponding list, since only a part of the desired
transaction has taken place.  

So the main ingredients of the model are: 
\emph{(i)}   {\bf Disorder}, since the initial money is distributed according to a Zipf's law and some parameters 
             characterizing the agents are randomly chosen, but the overall behavior is deterministic,
\emph{(ii)}  {\bf frustration}, because not all the individual wishes can be satisfied at the same time,
\emph{(iii)} {\bf delayed feedback}, involving the use of past values of the market price and volatility during the 
             decision formation,
\emph{(iv)}  {\bf phase transition} between the excess-demand and the excess-supply phases \cite{savit}.

\begin{figure}
\centerline{\psfig{file=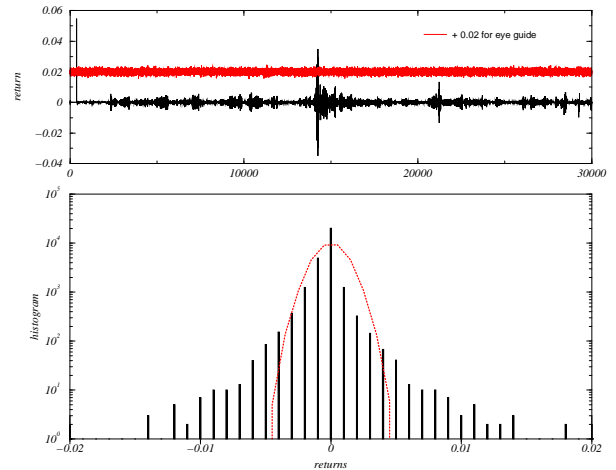,width=8cm,angle=270}}
\caption[]{\small\label{fig:fattails}
Upper panel: Price returns, random series shifted for eye guide. Lower panel: Returns distribution. The comparison
with a best-fitted normal distribution reveals the presence of \emph{fat tails}.}
\end{figure}

Fig. \ref{fig:serie} shows the result of a typical simulation on the market price and the corresponding amount
of exchanged shares. The model is able to reproduce all the typical features observed in empirical data. At the
beginning the price remains constant due to the \emph{IPO's phase}, namely the bank offers the shares to the
traders at the fixed price, the \emph{IPO price}. After that, one can see the typical pressure
made by agents who did not get enough shares during the initial public offer: The volumes are high and the price
tends to raise. Then, after a normal settlement, the price starts to oscillate with very low volumes: Traders with
shares do not want to sell because they hope to get more money if they wait a little bit more, agents without 
shares do not buy because the price is too high and there is no evidence of a trend on it.
Then oscillations become bigger and bigger and when the volumes are big as well, then a small crash occurs and the
price comes back to a more interesting value for potential buyers. As a consequence, volumes remain high and the 
price follows a so called rally period, followed again by a crash, maybe due to the fact that the bubble phase has
been too optimistic.

\begin{figure}
\centerline{\psfig{file=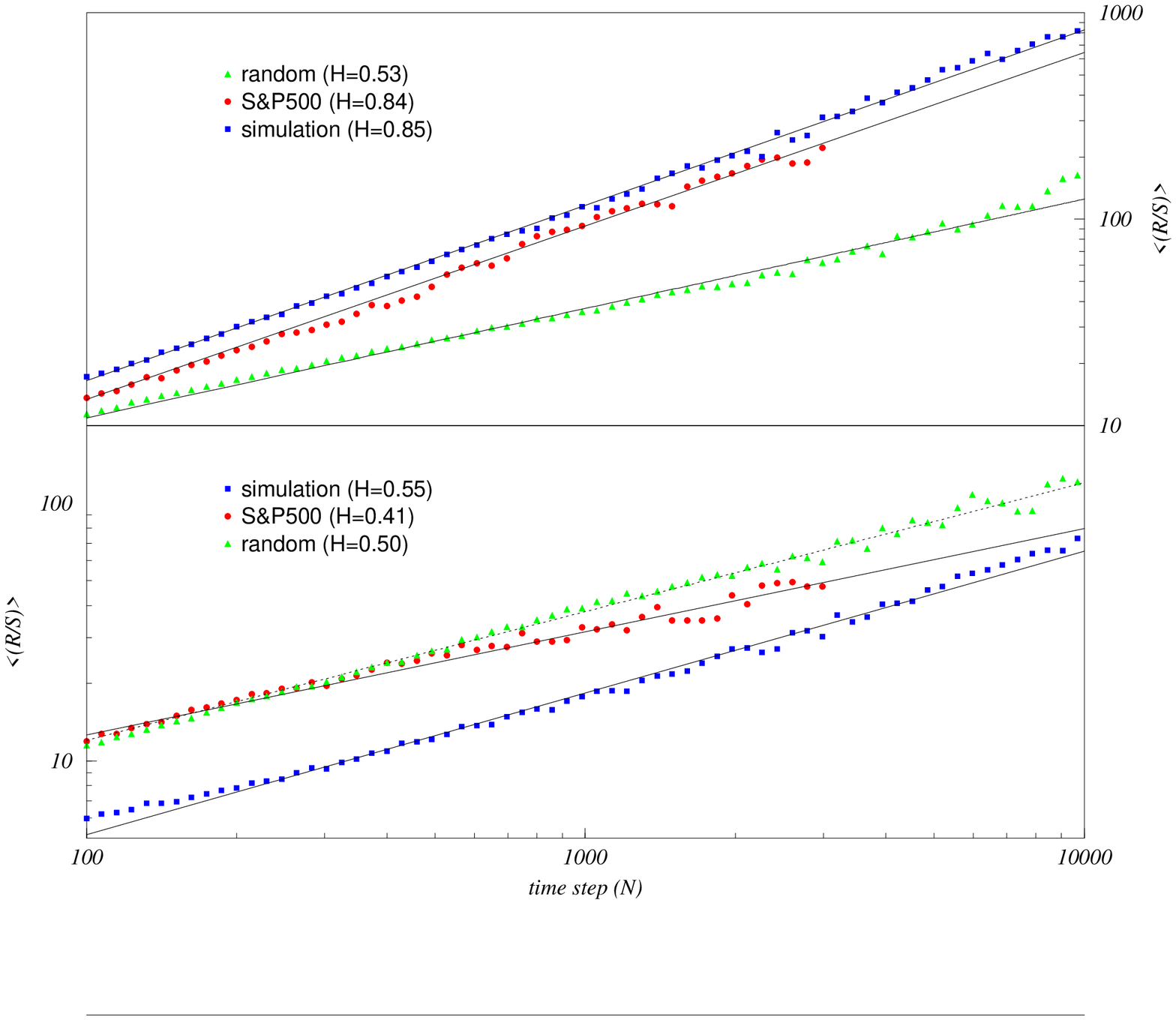,width=8cm,angle=270}}
\caption[]{\small\label{fig:hurst}
Typical snapshot from a simulation run. Estimation of the self-similarity parameter \emph{H}. Upper panel: Absolute
returns. Lower panel: Raw returns.}
\end{figure}

The model makes use of the following parameters: 
\emph{(i)} Number of traders {\bf N}: The number of agents involved in the process of buying and selling shares.
\emph{(ii)} Threshold {\bf T}: The critical age of the shares. After this time the trader may rise some doubts
                         whether it is worth waiting for the events. This value
						 is not constant for all the traders, but ranges among a decade in order to take into
						 account the differences between intraday speculators and long-time agents.
\emph{(iii)} Number of shares {\bf S} and IPO's price {\bf I}: Their product defines the initial value of the company. 
\emph{(iv)} Amount of money {\bf M} initially distributed among the traders. Inspired by \cite{marsili,zipf2} we 
            have decided to distribute the richness according to a Zipf's law: the 20 \% of the traders posses 
			around the 80 \% of {\bf M} and among the two groups this rule is applied again, recursively.
			In this way we are able to model the difference between a normal agent and an istitutional investor and the 
			different effect they produce when they decide to enter the market (responsible for the disorder). A minimal 
			value of money {\bf m} is provided to all the traders and added to the previous distribution.
\emph{(v)} Length of the past values' list {\bf MEM}: chartists look for trends and patterns in the 
           historical time series of the market price. 
\emph{(vi)} Unbalance of the book {\bf B}: This value takes into account how balanced are the sell 
            and the buy list. In case they are out of balance, then news and advertisements are generated.

As shown in Fig. \ref{fig:fattails}, the probability density function of the returns of our simulated stock shows a
strong leptokurtic nature. For comparison, the Gaussian with the same measured standard deviation is also reported. The
time series of returns exhibits a higher frequency of extreme events and clustering of volatility. This aspect becomes
clear thanks to Fig. \ref{fig:hurst}: When considering absolute returns as a measure of volatility, we see that the
transformed price data behave differently from their counterpart derived from the Gaussian distribution. The estimation
of the self-similarity parameter H (using the approach presented in \cite{peng}) reveals a strong persistence in 
volatility.

One comment about demand and supply. It had been a common sense in economics for a long time that demand and supply
balance automatically, however, it becomes evident that in reality such balances are hardly be realized for most of
popular commodities in our daily life \cite{takayasu2}. The important point is that demand is essentially a stochastic
variable because human action can never be predicted perfectly, hence the balance of demand and supply should also be
viewed in a probabilistic way. If demand and supply are balanced on average the probability of finding an 
arbitrarily chosen commodity on the shelves of a store should be 1/2, namely about half of the shelves should be
empty. Contrary to this theoretical estimation shelves in any department store or supermarket is nearly always
full of commodities. This clearly demonstrates that supply is much in excess in such stores.
In general the stochastic properties of demand and supply can be well characterized by a phase transition view which
consists of two phases: The excess-demand and excess-supply phases. It is a general property of a phase 
transition system that fluctuations are largest at the phase transition point, and this property also holds in 
this demand-supply system. In the case of markets of ordinary commodities, consumers and providers are independent
and the averaged supply and demand are generally not equal. The resulting price fluctuactions are generally slow
and small in such market because the system is out of the critical point.

\begin{figure}
\centerline{\psfig{file=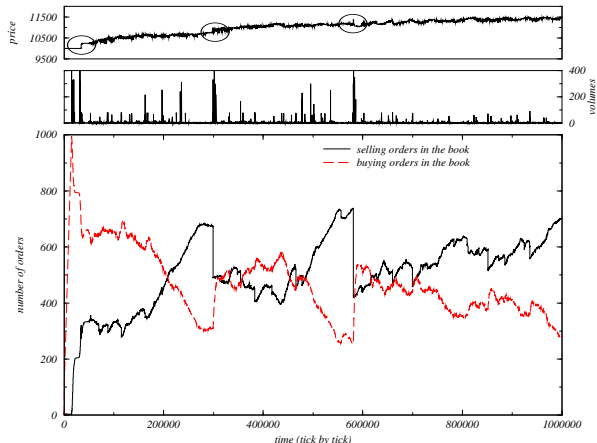,width=8cm,angle=270}}
\caption[]{\small\label{fig:sbilbook}
Simulation run with N = 1000.
Upper panel: Market price evolution on time. Middle panel: Volumes of exchanged shares on time.
Lower panel: Evolution of the book. Almost all the traders have placed
an order and are waiting. Note the symmetry of the two paths with respect to the half of the number 
of agents (N/2 = 500).}
\end{figure}

On the contrary in an open market of stocks or foreign exchanges, market is governed by speculative dealers who
frequently change their positions between buyers and sellers. It is shown that such speculative actions make demand
and supply balance automatically on average by changing the market price, as Fig. \ref{fig:sbilbook} clearly shows.
Contrary to \cite{bak} we do not need to impose that the number of the shares has to be half
of the number of traders in order to get a balance between demand and supply. The three circles in the upper panel 
indicate the most extreme events taking place in the price evolution: There are corresponding movements in the book
and in the volumes, since they are the reason for such a sudden variation. As the system is always at the critical
point the resulting price fluctuations are generally quick and large \cite{takayasu2}. This result is in agreement
with \cite{sornette}, where the authors present an analogy between large stock market crashes and critical points
with log-periodic correction to scaling: Complex systems often reveal more of their structure and organization
in highly stressed situations than in equilibrium.

Performing a correlation analysis on our simulations and comparing it to the results presented in \cite{mantegna}, 
we can associate a temporal scale to our tick by tick time: Since the autocorrelation function vanishes after 
approximately 20 ticks and the typical correlation length in financial time series is supposed to be around 20 
minutes, we can speculate that one of our tick corresponds to one real-life minute.

Therefore the involved time scales are the following:
\emph{(i)}   Total number of iterations, namely total number of ticks = $10^6$ ({\bf 10 years}),
\emph{(ii)}  threshold = 10000 ({\bf 1 month}),
\emph{(iii)} threshold variability in the range $[0.1,1]$. This gives a time variable from a minimum of {\bf 1 month} 
             to a maximum of {\bf 1 year} to have second thoughts. 

To summarize, we have proposed a model for the stockmarket which is able to reproduce the two main characteristics 
of empirical data, namely the correlated volatility and fat tails of the PDF of returns.
We have performed this task avoiding the use of different classes of agents and the artificial introduction of a
fundamental price. We just make use of realistic assumptions about the behavior of traders (limited amount of money,
limited time to liquidity, desired gain, maximum loss, inclination towards investment) and we tune the parameters 
through a real-life analysis of the trading activity. Disorder, frustration and the behavior at the critical point
do the rest. The model is intrinsecally a minority game in that agents in the minority are rewarded, i.e. they can
carry out their investment plan, and those in the majority punished, i.e. they have to wait with a pending order or
sell at an unfair price.

\end{multicols}

\end{document}